\documentclass[]{spie}  

\usepackage{amsmath,amsfonts,amssymb}
\usepackage{graphicx}
\usepackage[colorlinks=true, allcolors=blue]{hyperref}
\usepackage{xcolor}

\title{Early results from GRBAlpha and VZLUSAT-2}

\author[a]{Jakub \v{R}\'{\i}pa}
\author[b]{Andr\'as P\'al}
\author[c]{Masanori Ohno}
\author[a]{Norbert Werner}
\author[b]{L\'aszl\'o M\'esz\'aros}
\author[b]{Bal\'azs Cs\'ak}
\author[a]{Marianna Daf\v{c}\'{\i}kov\'a}
\author[d]{Vladim\'{\i}r D\'aniel}
\author[d]{Juraj Dud\'a\v{s}}
\author[e]{Marcel Frajt}
\author[f]{Peter Han\'ak}
\author[e]{J\'an Hudec}
\author[d]{Milan Junas}
\author[e]{Jakub Kapu\v{s}}
\author[g]{Miroslav Kasal}
\author[h]{Martin Koleda}
\author[h]{Robert Laszlo}
\author[f]{Pavol Lipovsk\'{y}}
\author[a]{Filip Münz}
\author[e]{Maksim Rezenov}
\author[f]{Miroslav \v{S}melko}
\author[d]{Petr Svoboda}
\author[c]{Hiromitsu Takahashi}
\author[i]{Martin Topinka}
\author[g]{Tom\'a\v{s} Urbanec}
\author[a]{Jean-Paul Breuer}
\author[j]{Teruaki Enoto}
\author[k]{Zsolt Frei}
\author[c]{Yasushi Fukazawa}
\author[a,k,l]{G\'abor Galg\'oczi}
\author[a]{Filip Hroch}
\author[m]{Yuto Ichinohe}
\author[b]{L\'aszl\'o L. Kiss}
\author[c]{Hiroto Matake}
\author[c]{Tsunefumi Mizuno}
\author[n]{Kazuhiro Nakazawa}
\author[o]{Hirokazu Odaka}
\author[c]{Helen Poon}
\author[p]{Nagomi Uchida}
\author[c]{Yuusuke Uchida}

\affil[a]{Department of Theoretical Physics and Astrophysics, Faculty of Science, Masaryk University,  Kotl\'a\v{r}sk\'a 267/2, Brno 611 37, Czech Republic}
\affil[b]{Konkoly Observatory, Research Centre for Astronomy and Earth Sciences, Budapest, Hungary}
\affil[c]{Hiroshima University, School of Science, Higashi-Hiroshima, Japan}
\affil[d]{Czech Aerospace Research Centre, Prague, Czech Republic}
\affil[e]{Spacemanic Ltd, Bratislava, Slovakia}
\affil[f]{Faculty of Aeronautics, Technical University of Ko\v{s}ice, Slovakia}
\affil[g]{Department of Radio Electronics, Faculty of Electrical Engineering and Communication, Brno University of Technology, Brno, Czech Republic}
\affil[h]{Needronix Ltd, Bratislava, Slovakia}
\affil[i]{INAF - Istituto di Astrofisica Spaziale e Fisica Cosmica, Via A. Corti 12, I-20133 Milano, Italy}
\affil[j]{The Hakubi Center for Advanced Research, Kyoto University, Kyoto, Japan}
\affil[k]{Institute of Physics, E\"otv\"os Lor\'and University, Budapest, Hungary}
\affil[l]{Wigner Research Centre, Budapest, Hungary}
\affil[m]{Department of Physics, Rikkyo University, Tokyo, Japan}
\affil[n]{Department of Physics, Nagoya University, Nagoya, Aichi, Japan}
\affil[o]{Department of Physics, The University of Tokyo, Bunkyo-ku, Tokyo, Japan}
\affil[p]{Institute of Space and Astronautical Science, Japan Aerospace Exploration Agency, Japan}

\authorinfo{Further author information: (Send correspondence to J. \v{R}\'{\i}pa)\\
J. \v{R}\'{\i}pa: E-mail: ripa.jakub@gmail.com}

\begin{document} 
\maketitle

\begin{abstract}
We present the detector performance and early science results from \textit{GRBAlpha}, a 1U CubeSat mission, which is a technological pathfinder to a future constellation of nanosatellites monitoring gamma-ray bursts (GRBs). \textit{GRBAlpha} was launched in March 2021 and operates on a 550\,km altitude sun-synchronous orbit.
The gamma-ray burst detector onboard \textit{GRBAlpha} consists of a $75\times75\times5\,{\rm mm}$ CsI(Tl) scintillator, read out by a dual-channel multi-pixel photon counter (MPPC) setup. It is sensitive in the $\sim 30-900$\,keV range. The main goal of \textit{GRBAlpha} is the in-orbit demonstration of the detector concept, verification of the detector's lifetime, and measurement of the background level on low-Earth orbit, including regions inside the outer Van Allen radiation belt and in the South Atlantic Anomaly.
\textit{GRBAlpha} has already detected five, both long and short, GRBs and two bursts were detected within a time-span of only 8 hours, proving that nanosatellites can be used for routine detection of gamma-ray transients. For one GRB, we were able to obtain a high resolution spectrum and compare it with measurements from the Swift satellite. We find that, due to the variable background, the time fraction of about 67\,\% of the low-Earth polar orbit is suitable for gamma-ray burst detection. One year after launch, the detector performance is good and the degradation of the MPPC photon counters remains at an acceptable level. The same detector system, but double in size, was launched in January 2022 on \textit{VZLUSAT-2} (3U CubeSat). It performs well and already detected three GRBs and two solar flares. Here, we present early results from this mission as well.
\end{abstract}

\keywords{gamma-rays; gamma-ray bursts; high-energy astrophysics; nano-satellites; instrumentation; detectors; scintillators; multi-pixel photon counter; low Earth orbit background}

\section{Introduction}
\label{sec:intro} 
We are living in revolutionary times, when nanosatellites are becoming complementary to large missions and can contribute to our understanding of the high-energy Universe. Several nanosatellite missions and small payloads are being planned, constructed or already launched, e.g. HERMES-TP/SP\cite{fiore2020}, GRID\cite{wen2019}, BurstCube\cite{perkins2020}, GALI\cite{rahin2020} and GTM\cite{chang2022}. 

The \textit{GRBAlpha}\cite{pal2020}\footnote{\url{https://grbalpha.konkoly.hu}} nanosatellite is a 1U CubeSat which caries an on-board gamma-ray detector capable of detecting hard X-ray / gamma-ray transient sources such as gamma-ray bursts (GRBs)\cite{kouveliotou2012}. It is a technological pathfinder for a future constellation of nanosatellites CAMELOT\cite{werner2018} monitoring and localizing GRBs using time-based cross-correlation technique\cite{ohno2018,ohno2020}.

The detector onboard \textit{GRBAlpha} consists of a $75\times75\times5\,{\rm mm}$ CsI(Tl) scintillator read out by eight silicon photomultipliers (SiPM), multi-pixel photon counters (MPPCs) S13360-3050 PE by Hamamatsu, arranged into two independent readout channels with four MPPCs per channel\cite{ohno2018,torigoe2019,pal2020}. The detector is sensitive in the $\sim30-900$\,keV range with the maximum effective area of $\sim 54$\,cm$^2$ at 100\,keV.

The satellite was launched from Baikonur by a Soyuz-2.1a rocket to a 550\,km altitude Sun-synchronous orbit (SSO) on March 22, 2021. The main goal of \textit{GRBAlpha} is the in-orbit demonstration of the detector concept, verification of the detector's lifetime, and measurement of the background level on low-Earth orbit (LEO), including regions inside the outer Van Allen radiation belt and in the South Atlantic Anomaly (SAA).

The main radio ground stations used for commanding the satellite and downloading the scientific data are: the ground station at the Brno University of Technology (Czech Republic); the Bankov ground station run by the Faculty of Aeronautics at the Technical University of Kosice (Slovak Republic) and the ground station at the Piszk\'estet\H o observatory (Hungary). The satellite is using amateur radio bands and the housekeeping and scientific data are available via the SatNOGS network\footnote{\url{https://satnogs.org}}.

\textit{VZLUSAT-2}\footnote{\url{https://www.vzlusat2.cz/en/}} is a 3U sized CubeSat, developed by the Czech Aerospace Research Centre as a technology mission,  with two Earth observing cameras as the primary payload. Among several secondary payloads (X-ray and particle sensors) are two perpendicularly placed GRB detectors. Each detector employs a $75\times75\times5\,{\rm mm}$ CsI(Tl) scintillator read out in the same manner as on {\it GRBAlpha}. The detectors are placed under customized solar panels with reduced copper layer for better X-ray transparency.

The satellite was launched from Cape Canaveral (USA) by a Falcon 9 rocket to a 550\,km altitude SSO on January 13, 2022. The main radio ground stations used for commanding or downloading the scientific data are: the station of the University of West Bohemia in Pilsen (Czech Republic); the station at the Piszk\'estet\H o observatory (Hungary) and the SatNOGS network.

In Sec.~\ref{sec:grbalpha-results}, we describe the main results of \textit{GRBAlpha}, in Sec.~\ref{sec:vzlusat2-results}, we present the early results from \textit{VZLUSAT-2}, and in Sec.~\ref{sec:summary}, we summarize the main conclusions.

\section{Results from \textit{GRBAlpha}}
\label{sec:grbalpha-results}

The \textit{GRBAlpha} nanosatellite has been successfully operating in orbit for more than one year. Below, we summarize the main results.

\begin{figure}[!ht]
\begin{center}
\resizebox{0.49\textwidth}{!}{\includegraphics{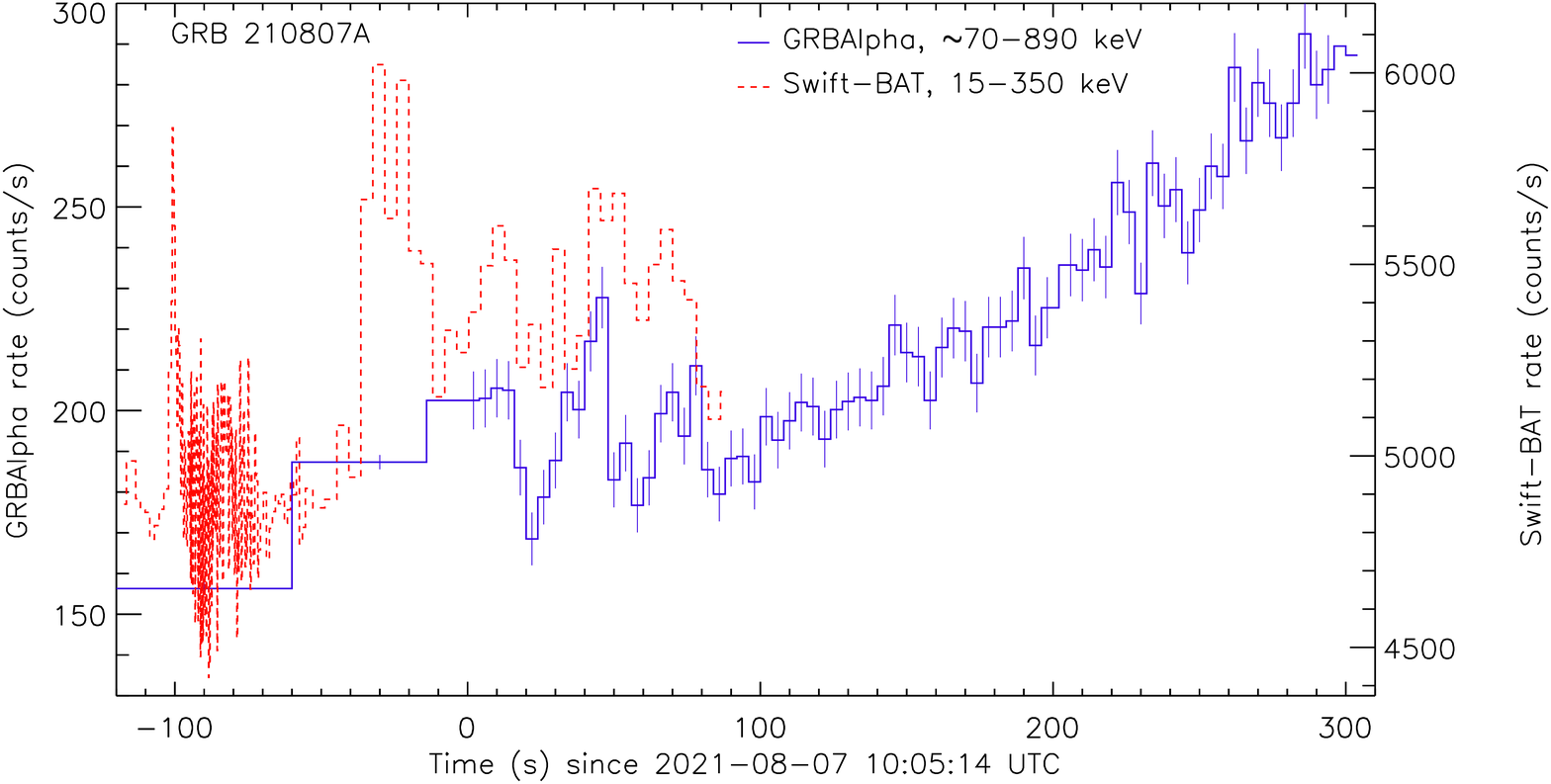}}\vspace*{2mm}
\resizebox{0.49\textwidth}{!}{\includegraphics{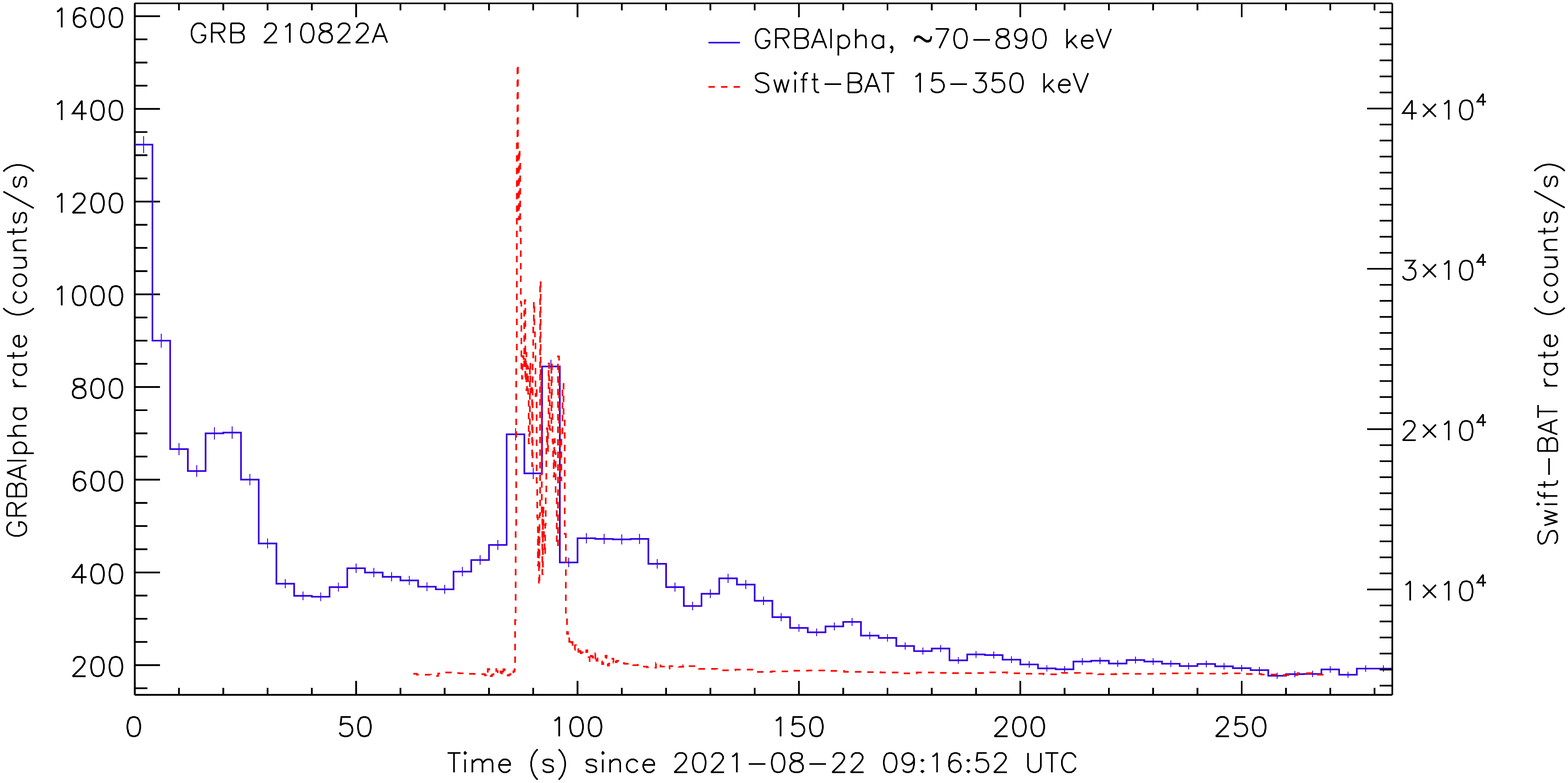}}\vspace*{2mm}
\resizebox{0.49\textwidth}{!}{\includegraphics{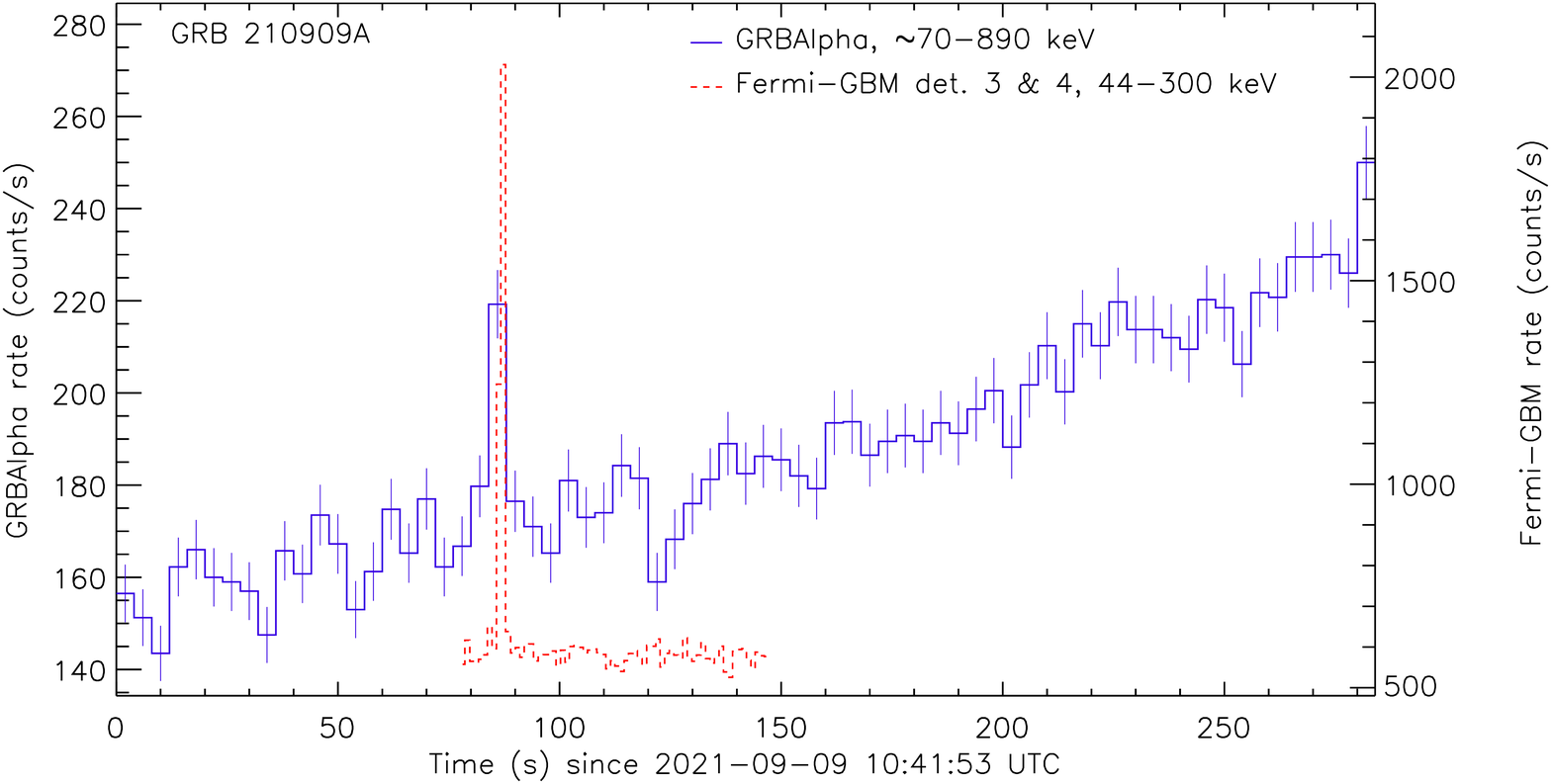}}\vspace*{2mm}
\resizebox{0.49\textwidth}{!}{\includegraphics{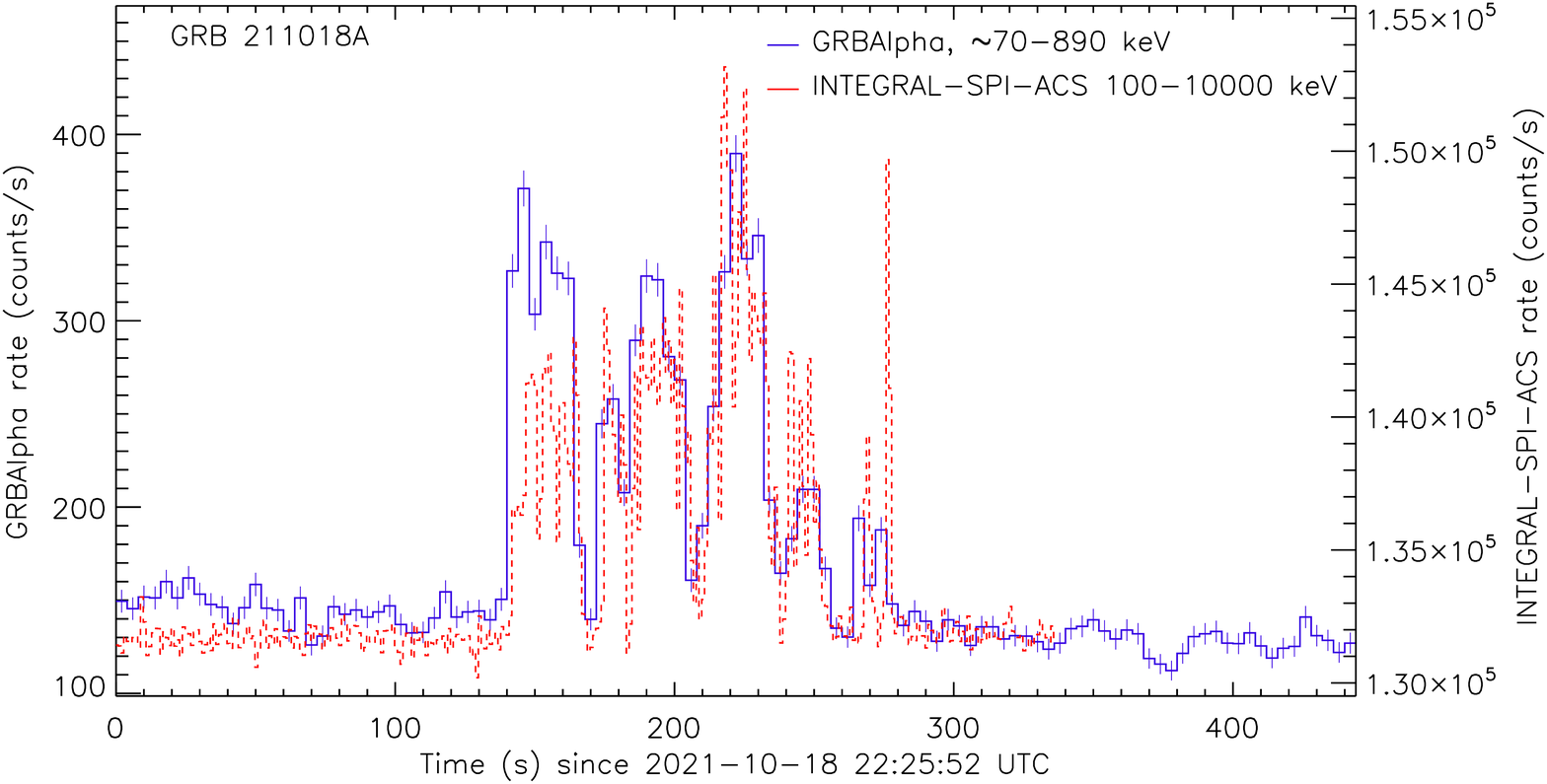}}\vspace*{2mm}
\resizebox{0.49\textwidth}{!}{\includegraphics{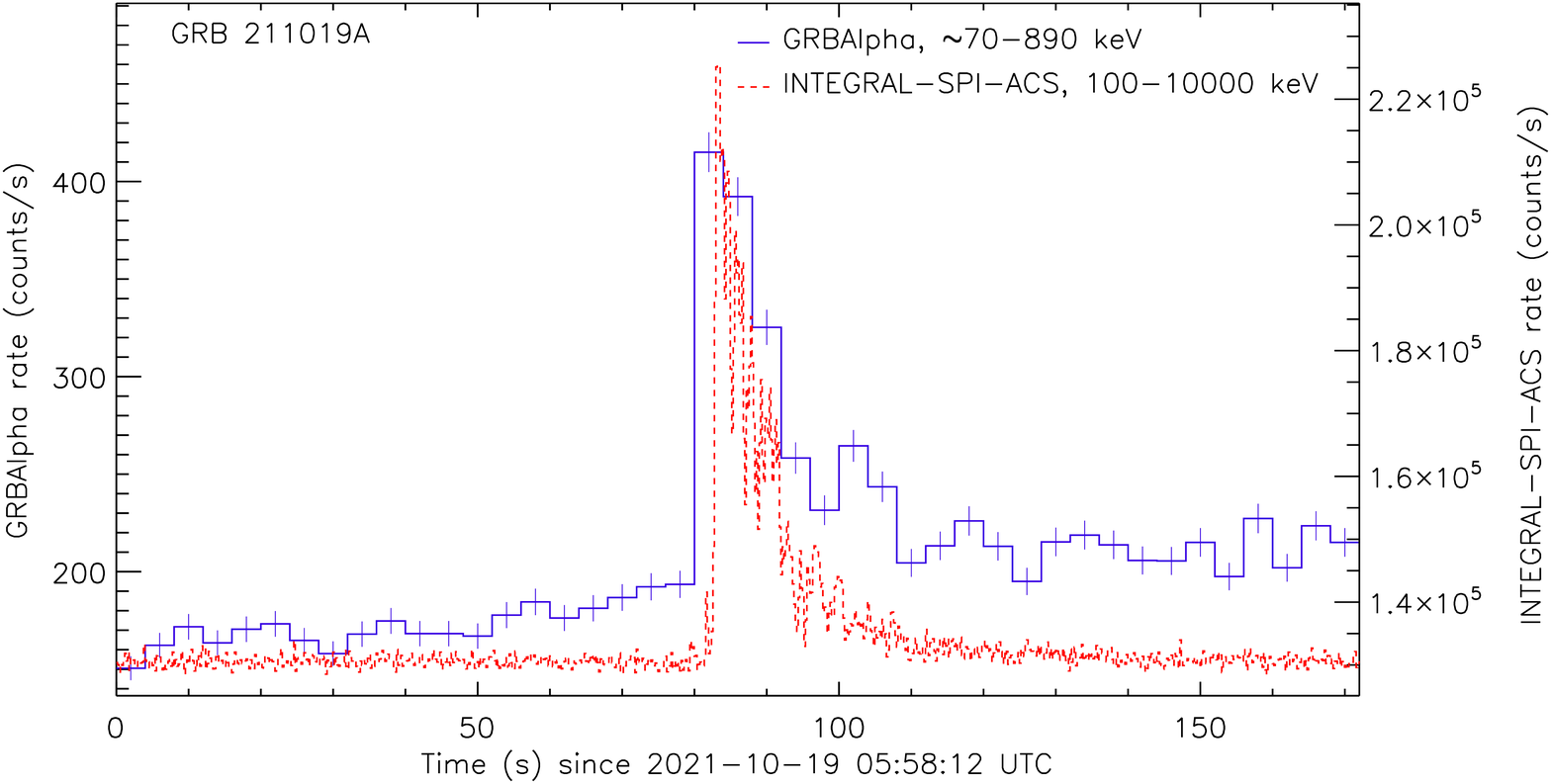}}\vspace*{2mm}
\caption{Five gamma-ray bursts (GRB 210807A, GRB 210822A, GRB 210909A, GRB 211018A and GRB 211019A) detected so far by \textit{GRBAlpha} over-plotted with raw light curves measured by other missions.}
\label{fig:grbalpha-grbs}
\end{center}
\end{figure}

\subsection{GRB Detections}
\label{sec:grbalpha-events}

\textit{GRBAlpha} has so far detected five GRBs, which were simultaneously observed by other satellites. The current version of the onboard payload firmware does not contain a trigger algorithm which would autonomously detect gamma-ray transients because the focus of this mission was mainly to verify the detector concept. The binned count data are recorded on board. The number of spectral bins (up to 256) as well the binning in time is adjustable by a command from ground. Once a new nominal measurement session is started then up to $\sim8$ hours of continuous count rate curves with 4\,s temporal resolution and with 4 energy bands are recorded. On ground the GCN\cite{barthelmy1998}\footnote{\url{https://gcn.gsfc.nasa.gov/gcn_main.html}} notices with GRB alerts from \textit{Fermi}/GBM, \textit{Swift}/BAT, \textit{INTEGRAL}/SPI-ACS, \textit{AGILE}/MCAL, \textit{Wind}-KONUS, \textit{CALET}/CGBM, \textit{MAXI} and other public GRB trigger lists, e.g. ASTROSAT/CZTI GRB archive\footnote{\url{http://astrosat.iucaa.in/czti/?q=grb}} and Gamma-ray Urgent Archiver for Novel Opportunities (GUANO)\cite{Tohuvavohu2020}\footnote{\url{https://www.swift.psu.edu/guano}} are used to decide which intervals of data recorded by \textit{GRBAlpha} are worth downloading to search for gamma-ray transients. The five observed GRBs are plotted in Fig.~\ref{fig:grbalpha-grbs}.

The first detection was GRB 210807A (GCN 30624)\cite{GCN30624}, which was simultaneously observed by \textit{Swift}/BAT. This was the first time that a GRB was observed by a 1U CubeSat. It was a long GRB which was detected with a signal-to-noise ratio (SNR) of 8. The data acquisition by \textit{GRBAlpha} started with $2\times 60$\,s long exposures and with full spectral resolution of 256 energy channels. Then it was followed by the nominal data acquisition with 4\,s temporal resolution in 4 energy bands. By chance the data acquisition with high spectral resolution started when this long GRB was ongoing and therefore we were able to perform a joint spectral analysis with the \textit{Swift}/BAT data. The spectrum measured by \textit{GRBAlpha} showed a good agreement with the one measured by \textit{Swift}/BAT. The best fit power-law has a slope of $\Gamma=0.90\pm0.1$ and a flux of $1.2\times10^{-7}$\,erg\,cm$^{-2}$\,s$^{-1}$ in the range of $50-300$\,keV.

The other detected GRBs by \textit{GRBAlpha} were: long GRB 210822A (GCN 30697)\cite{GCN30697} with SNR$\approx45$, short GRB 210909A (GCN 30840)\cite{GCN30840} with SNR=9, long GRB 211018A (GCN 30945)\cite{GCN30945} with SNR=46 and long GRB 211019A (GCN 30946)\cite{GCN30946} with SNR=39. 

GRB 210822A has a measured redshift of z=1.736 (GCN 30692)\cite{GCN30692}, which for the concordance cosmological model ($\Lambda$CDM with H$_0=69.6$\,km\,s$^{-1}$\,Mpc$^{-1}$, $\Omega_\mathrm{M}=0.29$, $\Omega_\mathrm{0}=0.71$) gives a light travel time of 9.9\,Gyr. GRB 211018A and GRB 211019A were detected within only 8 hours, demonstrating that nanosatellites can host payloads sensitive enough to routinely detect GRBs.

\begin{figure}[!ht]
\begin{center}
\resizebox{0.49\textwidth}{!}{\includegraphics{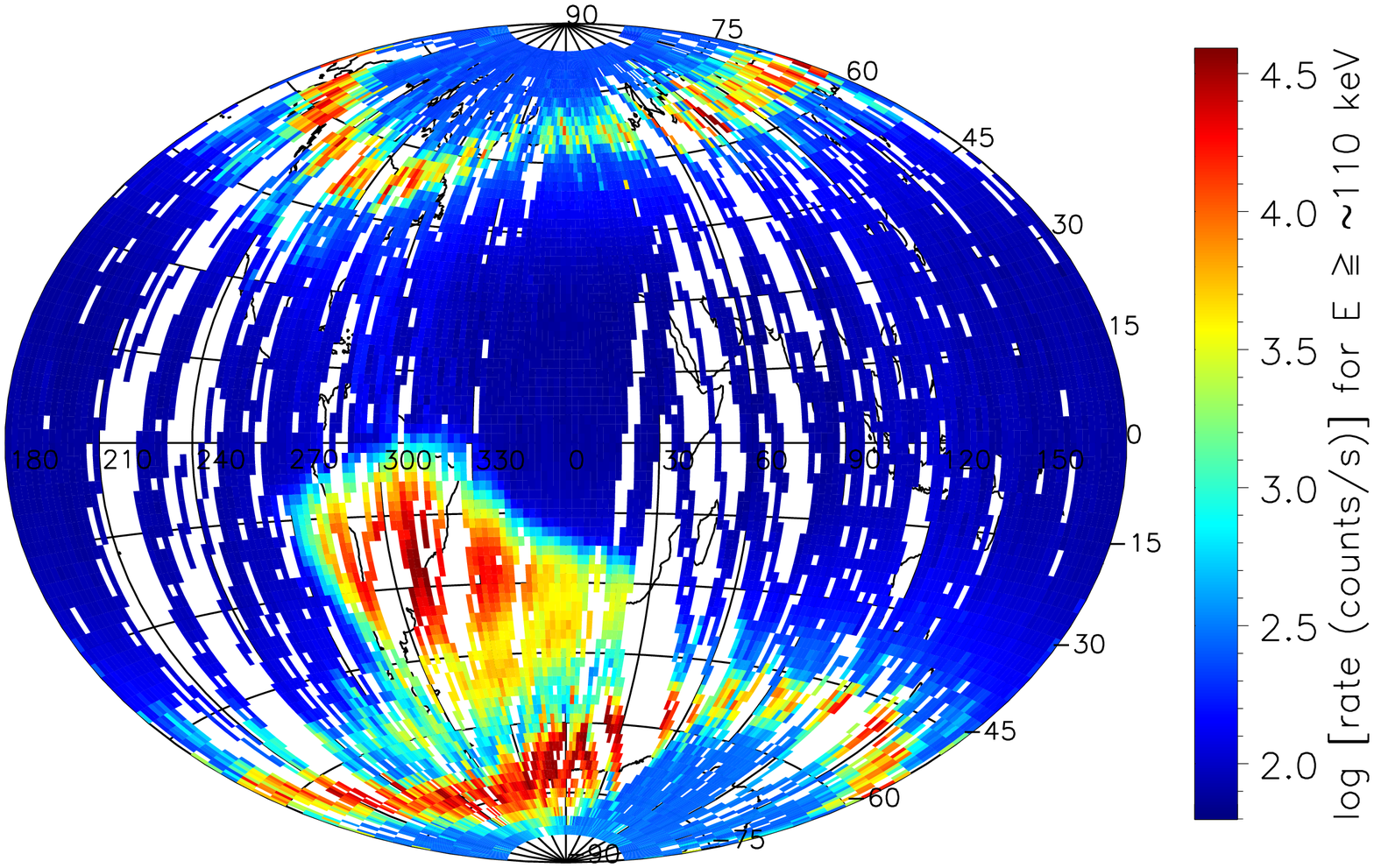}}\vspace*{2mm}
\resizebox{0.49\textwidth}{!}{\includegraphics{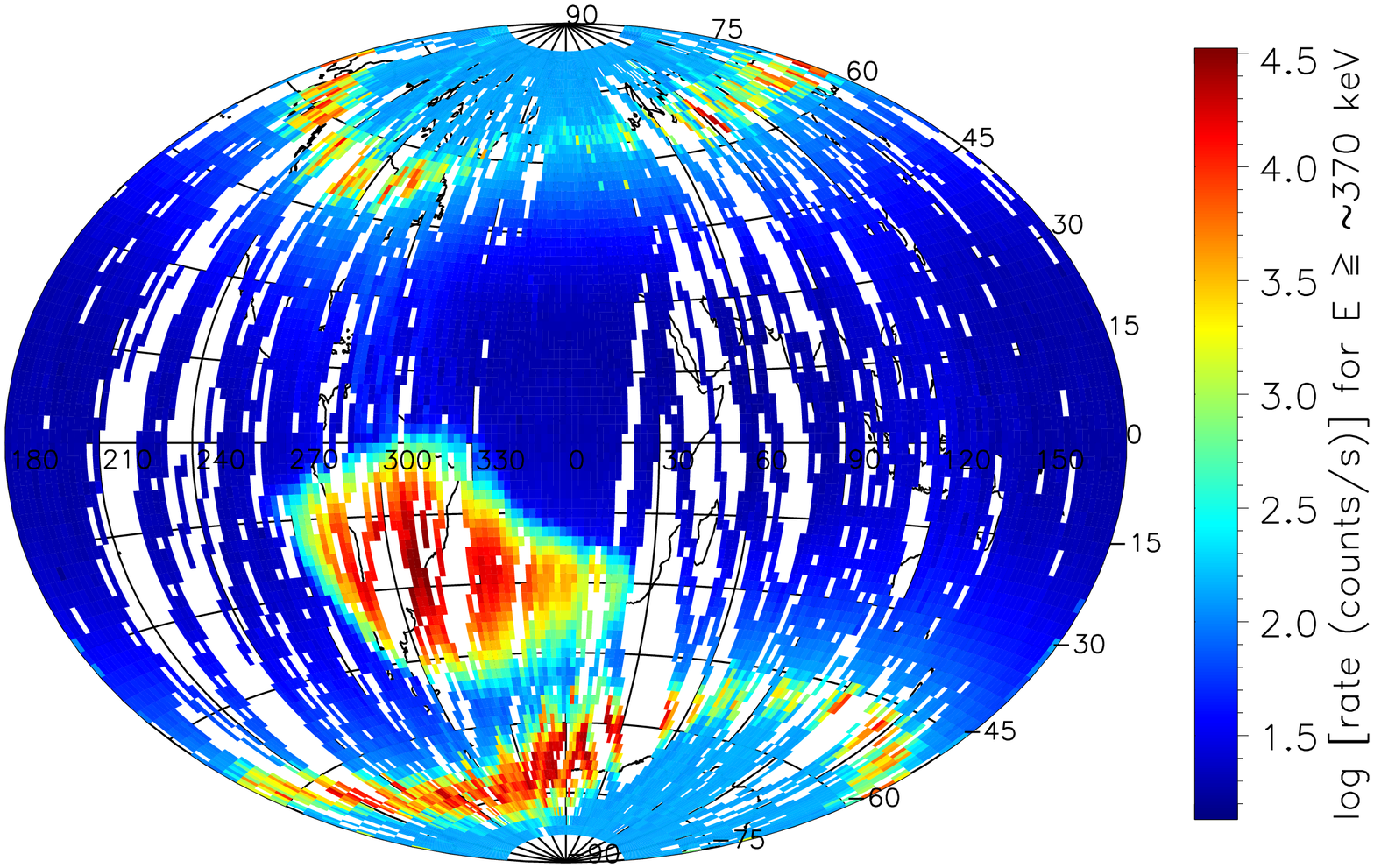}}\vspace*{2mm}
\caption{The background map at low Earth orbit (530-590\,km altitude) as measured by \textit{GRBAlpha} for $E\gtrsim 110$\,keV (left) and $E\gtrsim 370$\,keV (right), respectively. The count rate is averaged in a $2^\circ\times2^\circ$ pixels grid.}
\label{fig:grbalpha-bkg}
\end{center}
\end{figure}

\subsection{Low Earth Orbit Background Monitoring}
\label{sec:grbalpha-background}

The \textit{GRBAlpha} data also provide valuable information about the background environment at LEO which has several components, e.g. cosmic X-ray background (CXB), cosmic ray particles, geomagnetically trapped particles and albedo (secondary) emission produced in the Earth's atmosphere\cite{campana2013,galgoczi2021,campana2022}. The scintillator based gamma-ray detector is also sensitive to the energetic protons and electrons present in the Van Allen radiation belts (SAA and polar regions) \cite{ripa2020}. The advantage of our detector design, which employs MPPC sensors is that it operates at relatively low bias voltage and it can also safely measure inside the extreme radiation conditions of the SAA and polar regions.

\textit{GRBAlpha} is collecting data to create an average background map at LEO, which will help to estimate the expected duty cycle and to develop the rate trigger algorithms for future GRB missions. For the rate trigger, it is important to know when it should be enabled/disabled in order to avoid an extensive number of false detections. Fig.~\ref{fig:grbalpha-bkg} shows the averaged background maps in two energy bands collected from measurements with temporal resolution of 1, 4, and 15\,s.

\begin{figure}[!ht]
\begin{center}
\resizebox{0.8\textwidth}{!}{\includegraphics{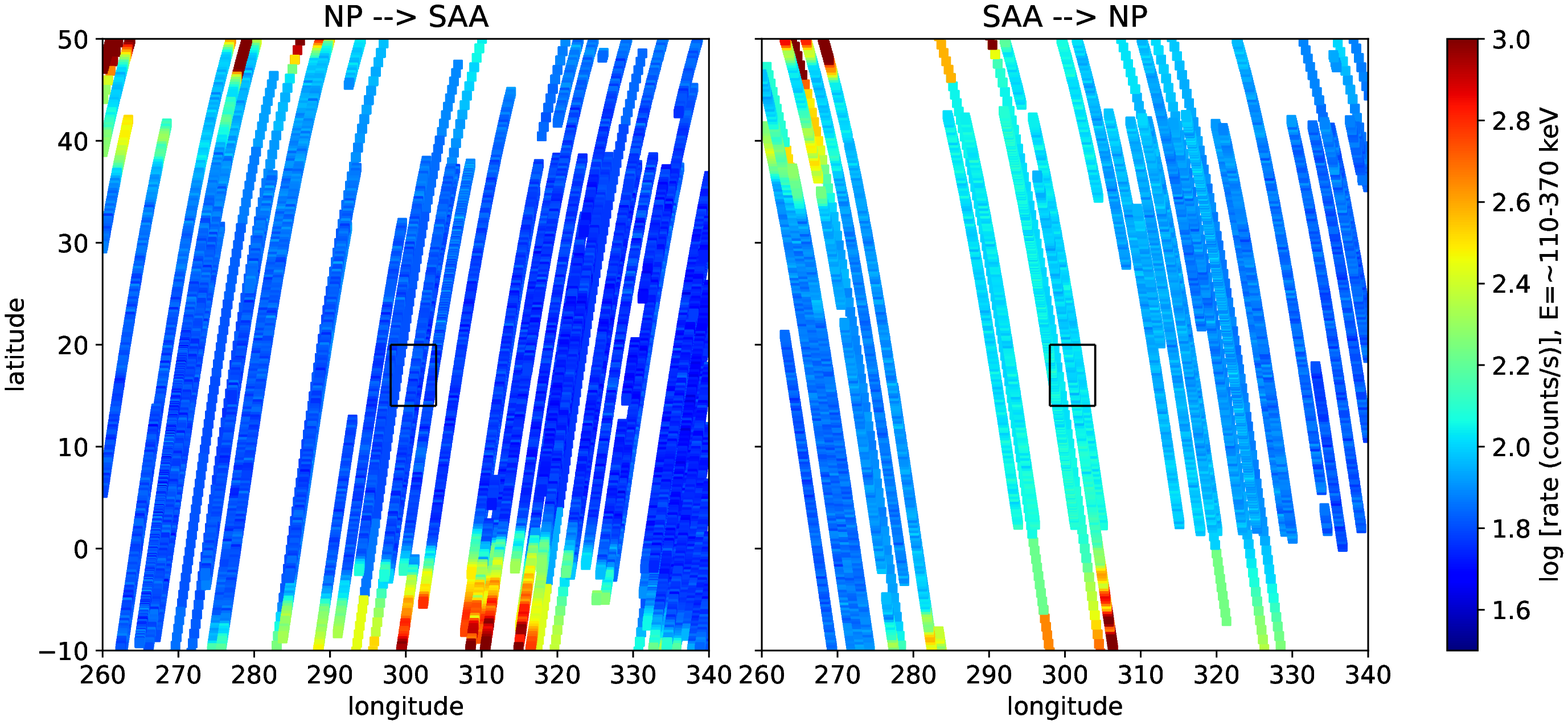}}\vspace*{2mm}
\resizebox{0.8\textwidth}{!}{\includegraphics{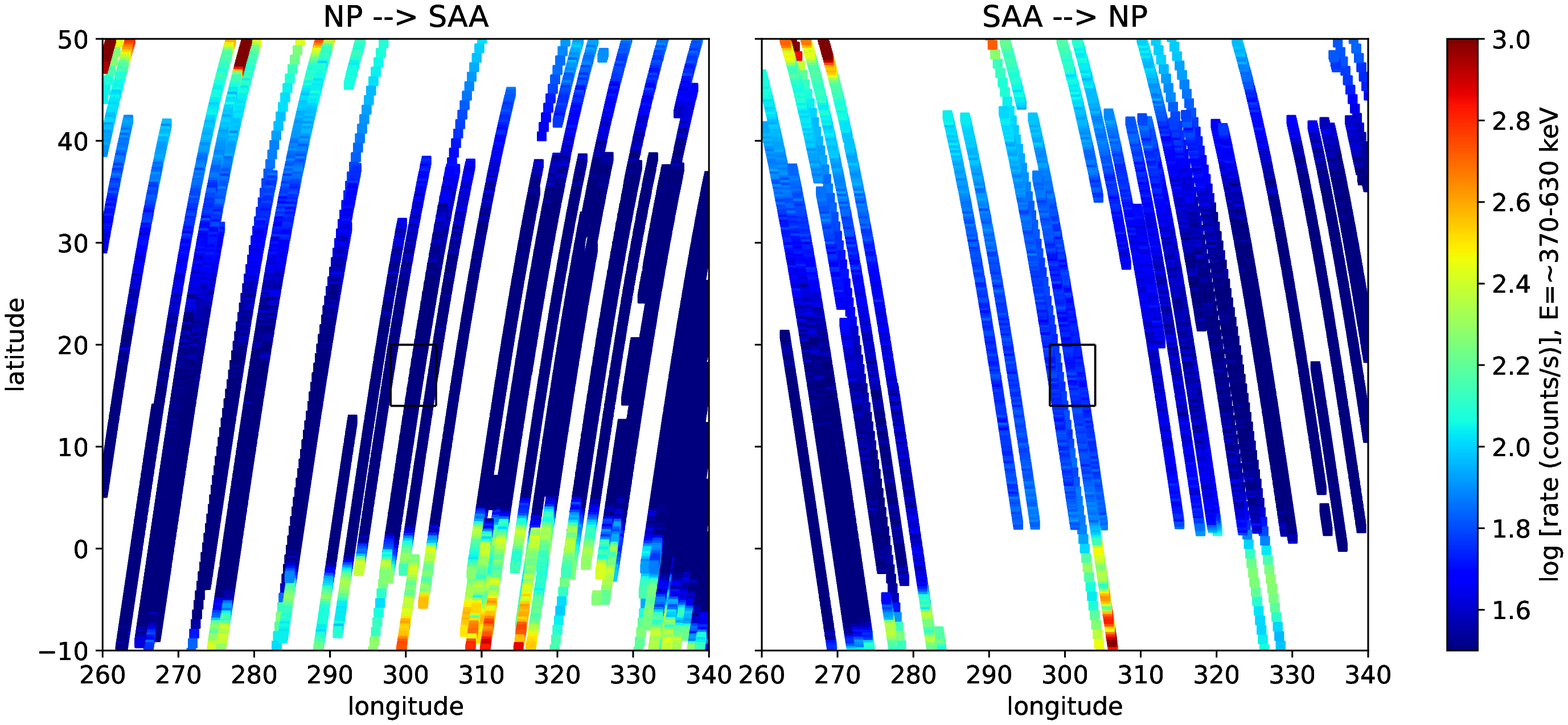}}\vspace*{2mm}
\caption{The maps of southward (left panels) and northward (right panels) ground tracks of \textit{GRBAlpha}, zoomed into the region between the South Atlantic Anomaly (SAA) and the north polar region showing the short-term activation background when the satellite passes SAA and moves northward. The maps show the count rate for the energy band of $E\approx 110-370$\,keV (upper panels) and $E\approx 370-630$\,keV (lower panels). The square in the middle of each panel marks the region in which the average count rate was calculated.}
\label{fig:grbalpha-activation}
\end{center}
\end{figure}

We noticed an interesting feature in the maps while comparing the measured background levels when the satellite travels from the north southward to the SAA and when it travels northward first passing the SAA. It is displayed in Fig.~\ref{fig:grbalpha-activation}. The measured mean background to the north of the SAA is significantly higher when the satellite passes SAA first. We calculated the mean background for the region of lon = 298$^\circ$-304$^\circ$ and lat = 14$^\circ$-20$^\circ$. The mean count rates are for the northward / southward passes: $105.3 \pm 0.7$ / $61.2 \pm 0.5$ (for E\,$\approx110-370$\,keV) and $55.9 \pm 0.4$ / $25.3 \pm 0.3$ (for E\,$\approx370-630$\,keV). This is likely due to the short-term radioactivation background induced by geomagnetically trapped protons in the SAA \cite{gruber1989,odaka2018,galgoczi2020}.

\subsection{Long-term Detector Performance}
\label{sec:grbalpha-detector}

\begin{figure}[!t]
\begin{center}
\resizebox{0.5\textwidth}{!}{\includegraphics[trim={0 0 20mm 13mm},clip]{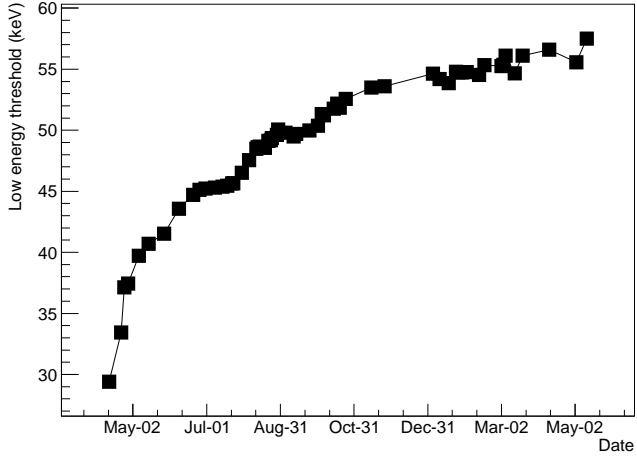}}\vspace*{2mm}
\caption{Increase of the low-energy threshold of the detector on board of \textit{GRBAlpha} plotted from the middle of April 2021 to the middle of May 2022. The shift of the low-energy threshold is caused by the increased noise in MPPCs a result of their gradual degradation due to the irradiation by geomagnetically trapped protons during SAA passages.}
\label{fig:grbalpha-threshold}
\end{center}
\end{figure}

One of the objectives of {\it GRBAlpha} was to study the aging of S13360-3050 PE MPPCs by Hamamatsu at LEO. As the MPPCs are exposed to geomagnetically trapped protons in the SAA they experience radiation damage which increases dark current and thus noise \cite{hirade2021}. In order to reduce the radiation damage the edge of the detector, where the MPPCs are placed, is shielded by 2.5\,mm antimony-lead alloy PbSb4. This shield together with a 1.0\,mm Al detector casing provides significant reduction of the total ionizing dose due to geomagnetically trapped and solar protons \cite{ripa2019}. The PbSb4 shield was designed based on the Multi-Layered Shielding Simulation tool (MULASSIS) \cite{Lei2002} implemented in SPENVIS\footnote{\url{www.spenvis.oma.be}}. Fig.~\ref{fig:grbalpha-threshold} demonstrates the shift of the low-energy threshold over the months since the launch. Before the launch the low-energy threshold was $\sim10$\,keV. Three weeks after the launch the threshold was $\sim29$\,keV and one year later it was $\sim56$\,keV. One year after the launch, the detector performance is good and the degradation of MPPCs remains at an acceptable level.

\begin{figure}[!t]
\begin{center}
\resizebox{0.7\textwidth}{!}{\includegraphics{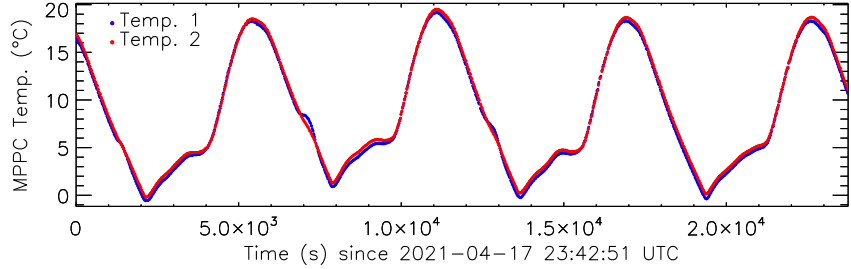}}\vspace*{2mm}
\caption{An example of the temperature variation measured over four orbits by two thermometers placed on the MPPC board inside the detector casing on \textit{GRBAlpha}.}
\label{fig:grbalpha-temp}
\end{center}
\end{figure}

Fig.~\ref{fig:grbalpha-temp} shows an example of the temperature variation measured by thermometers placed on the MPPC board of the detector on board of \textit{GRBAlpha}.

\section{Results from VZLUSAT-2}
\label{sec:vzlusat2-results}
The \textit{VZLUSAT-2} nanosatellite finished the commissioning phase after the launch and it is now nominally operating. Below are summarized the main results.

\subsection{GRB and Solar Flare Detections}
\label{sec:vzlusat2-events}

The GRB payload on \textit{VZLUSAT-2} is another technology demonstration and a rate trigger, which would automatically detect GRBs, is not yet implemented in the onboard software. GCN notices and other public GRB trigger lists are used to determine which data sections will be downloaded. There are two GRB detector units on board marked as unit no. 0 and 1. Nominal measurements are done with 1\,s temporal resolution in 4 energy bands.

\begin{figure}[!t]
\begin{center}
\resizebox{0.49\textwidth}{!}{\includegraphics{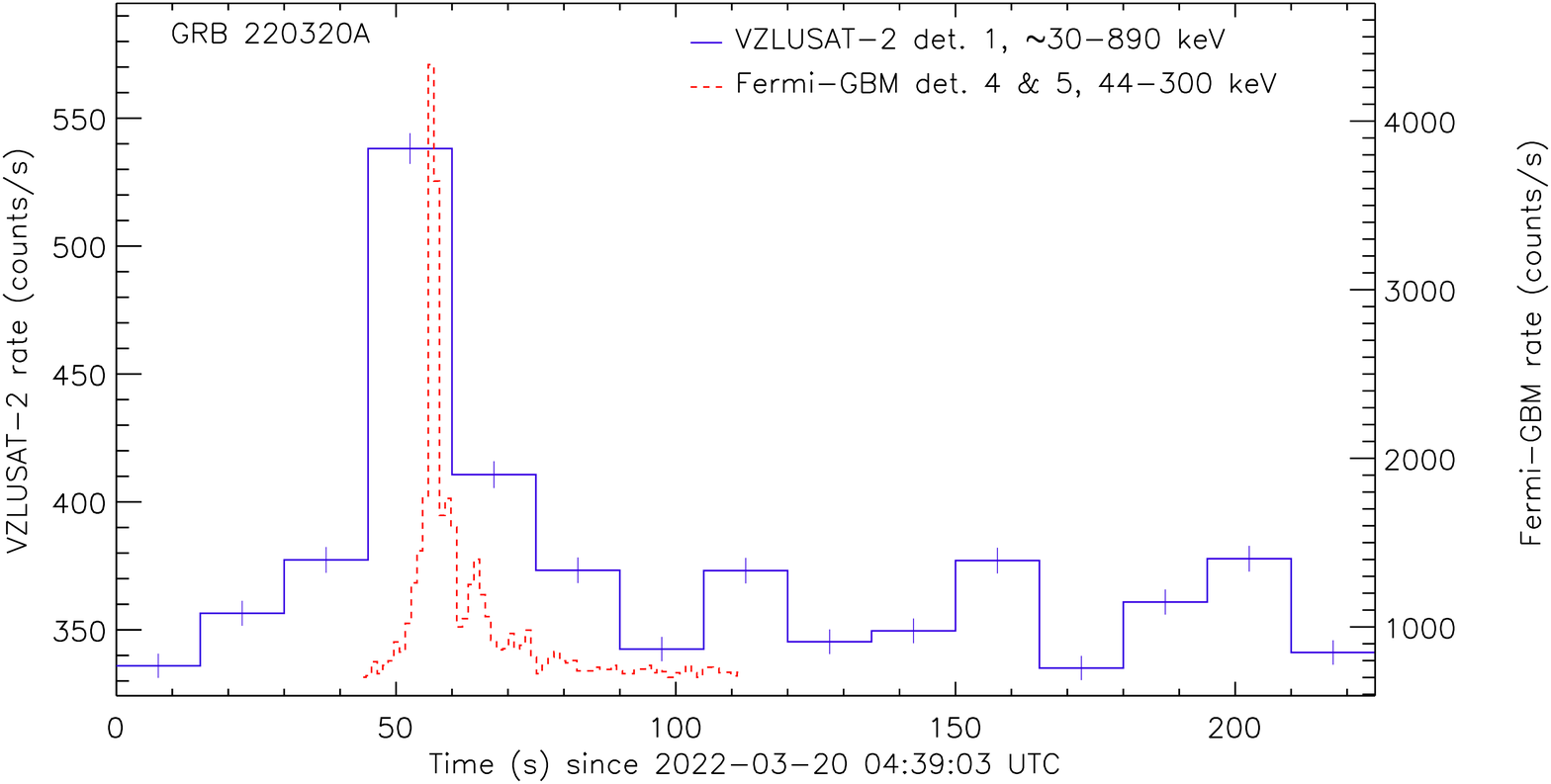}}\vspace*{2mm}
\resizebox{0.49\textwidth}{!}{\includegraphics{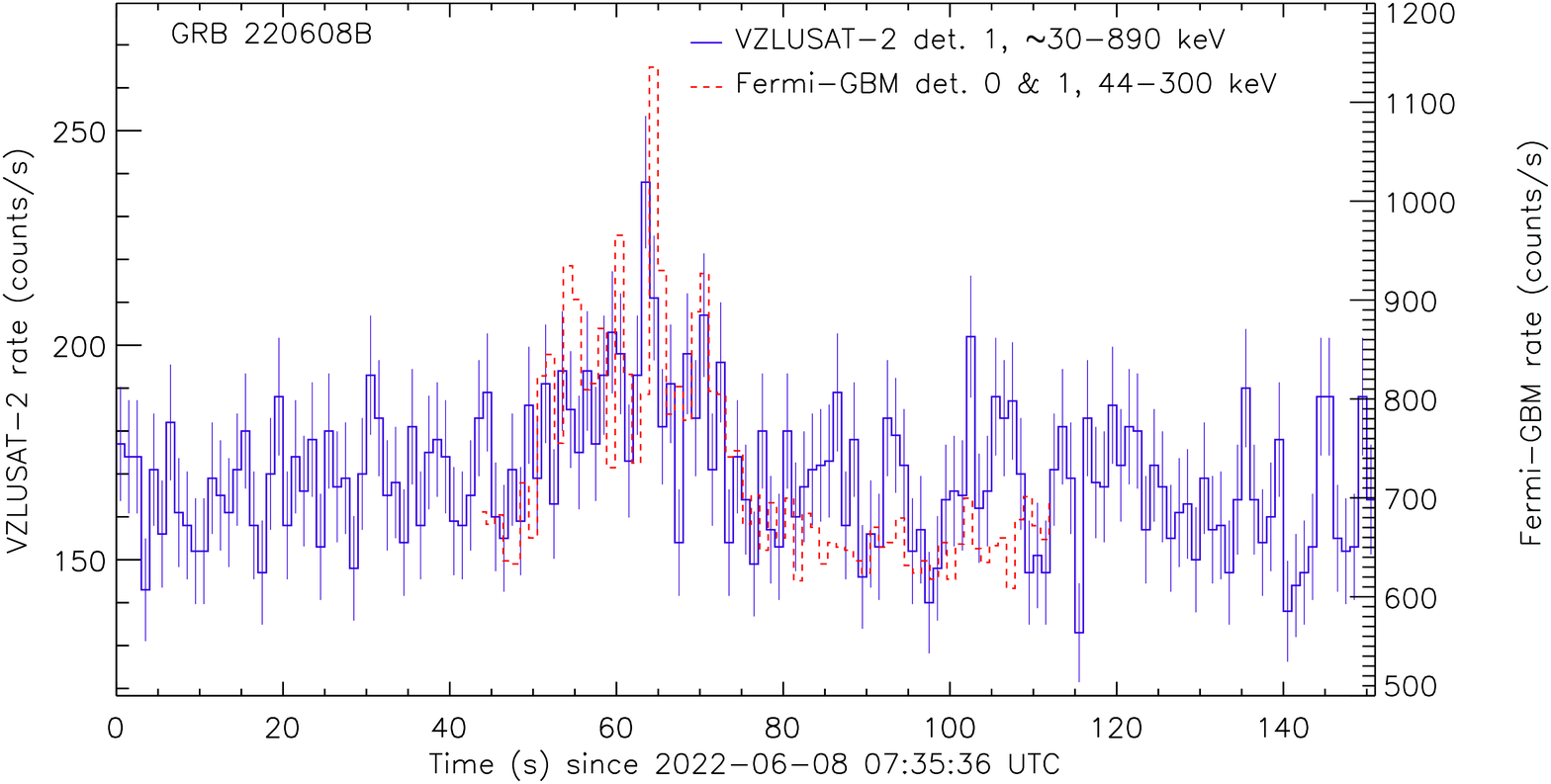}}\vspace*{2mm}
\resizebox{0.49\textwidth}{!}{\includegraphics{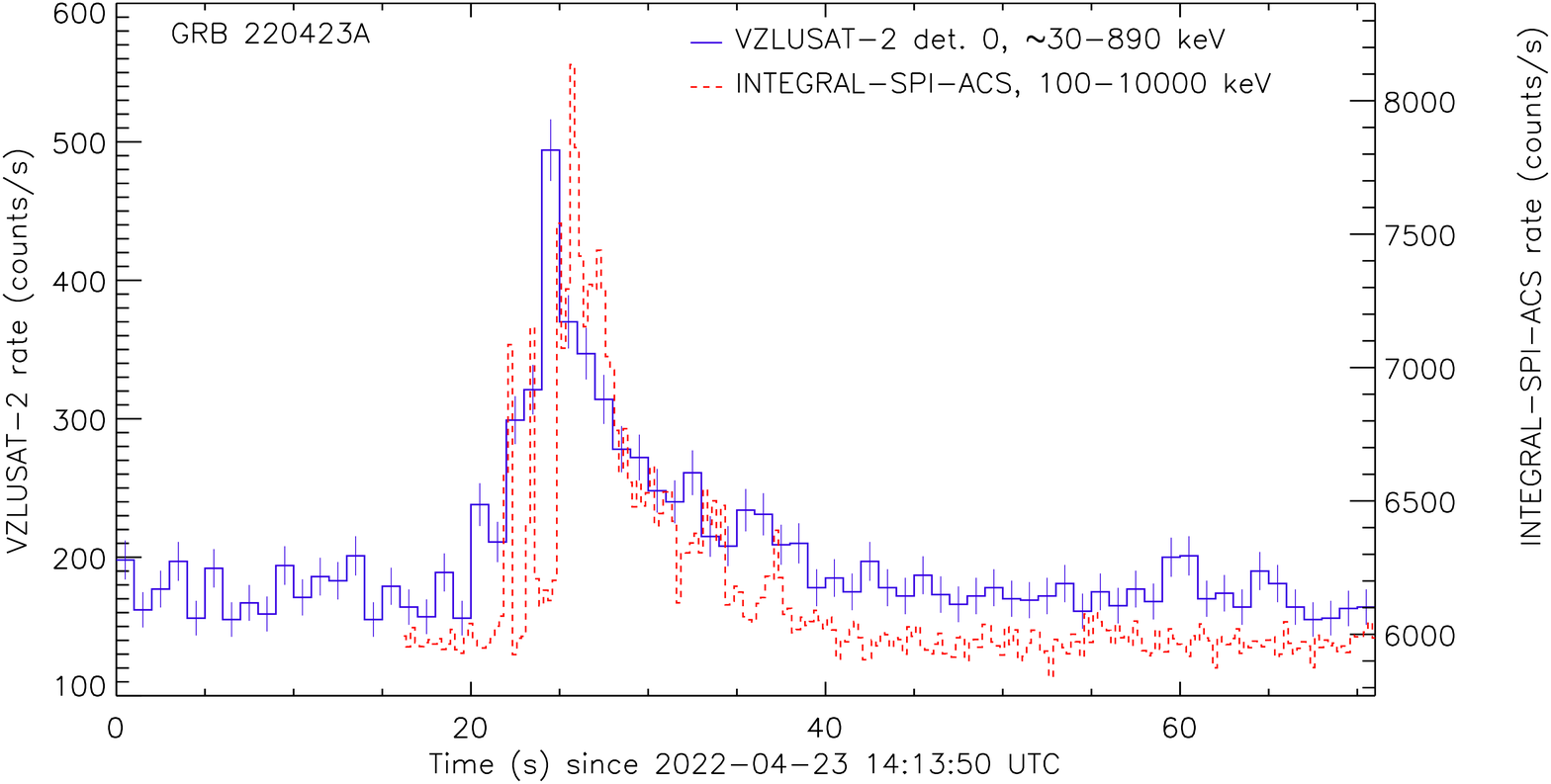}}\vspace*{2mm}
\resizebox{0.49\textwidth}{!}{\includegraphics{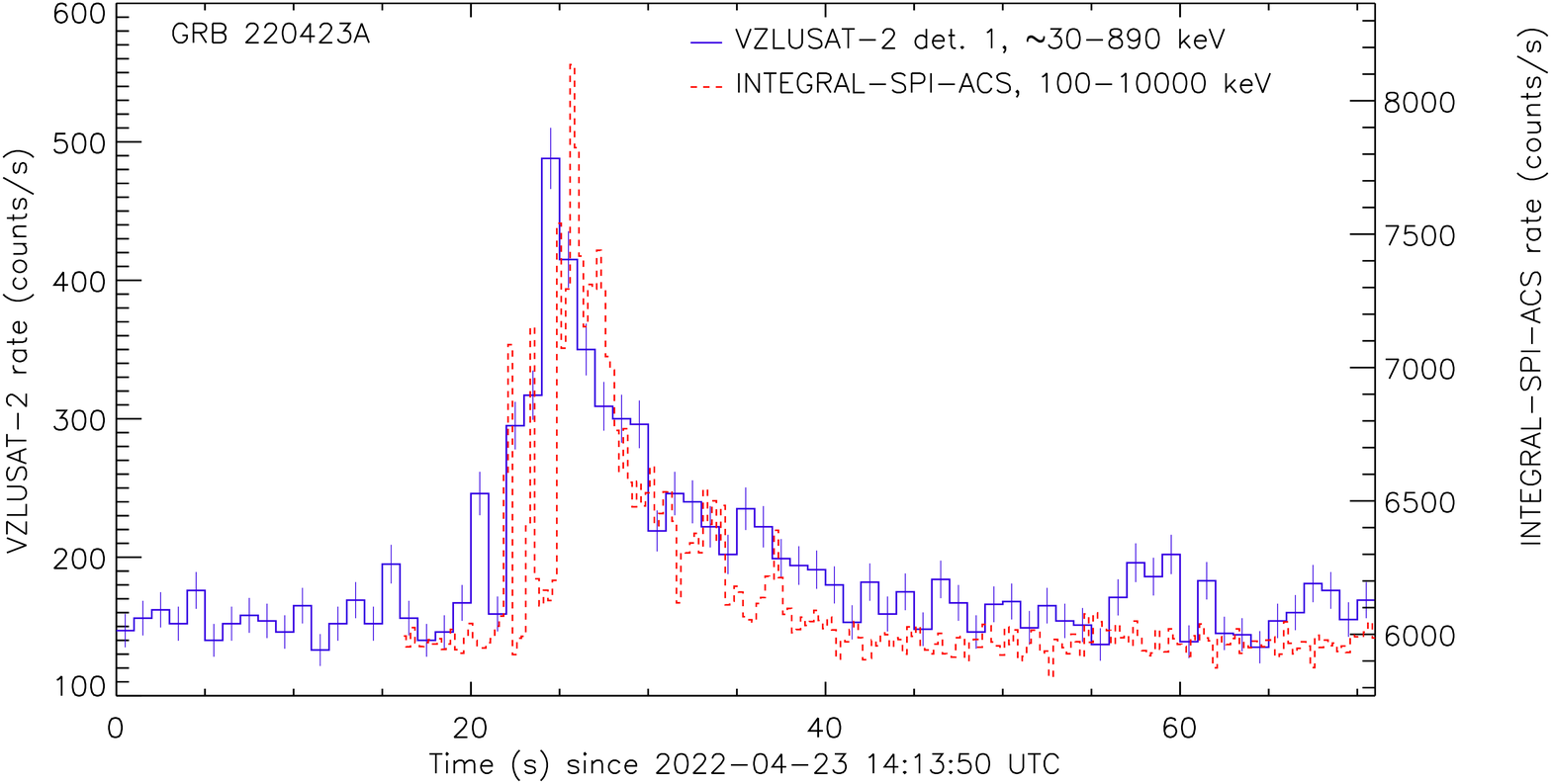}}\vspace*{2mm}
\caption{Three gamma-ray bursts (GRB 220608B, GRB 220423A and GRB220320A) detected so far by the GRB payload on \textit{VZLUSAT-2} over-plotted with raw light curves measured by other missions. The bottom panels show the same burst observed by the two perpendicularly placed detector units on \textit{VZLUSAT-2}.}
\label{fig:vzlusat2-grbs}
\end{center}
\end{figure}

The GRB payload on \textit{VZLUSAT-2} has, at the time of writing, detected three long GRBs:
GRB 220320A (GCN 32196)\cite{GCN32196} with SNR=6, GRB 220423A (GCN 31965)\cite{GCN31965} with SNR=28 and GRB 220608B (GCN 31803)\cite{GCN31803} with SNR=46. The events detected by \textit{VZLUSAT-2} are plotted in Fig.~\ref{fig:vzlusat2-grbs}.

\begin{figure}[!t]
\begin{center}
\resizebox{0.49\textwidth}{!}{\includegraphics{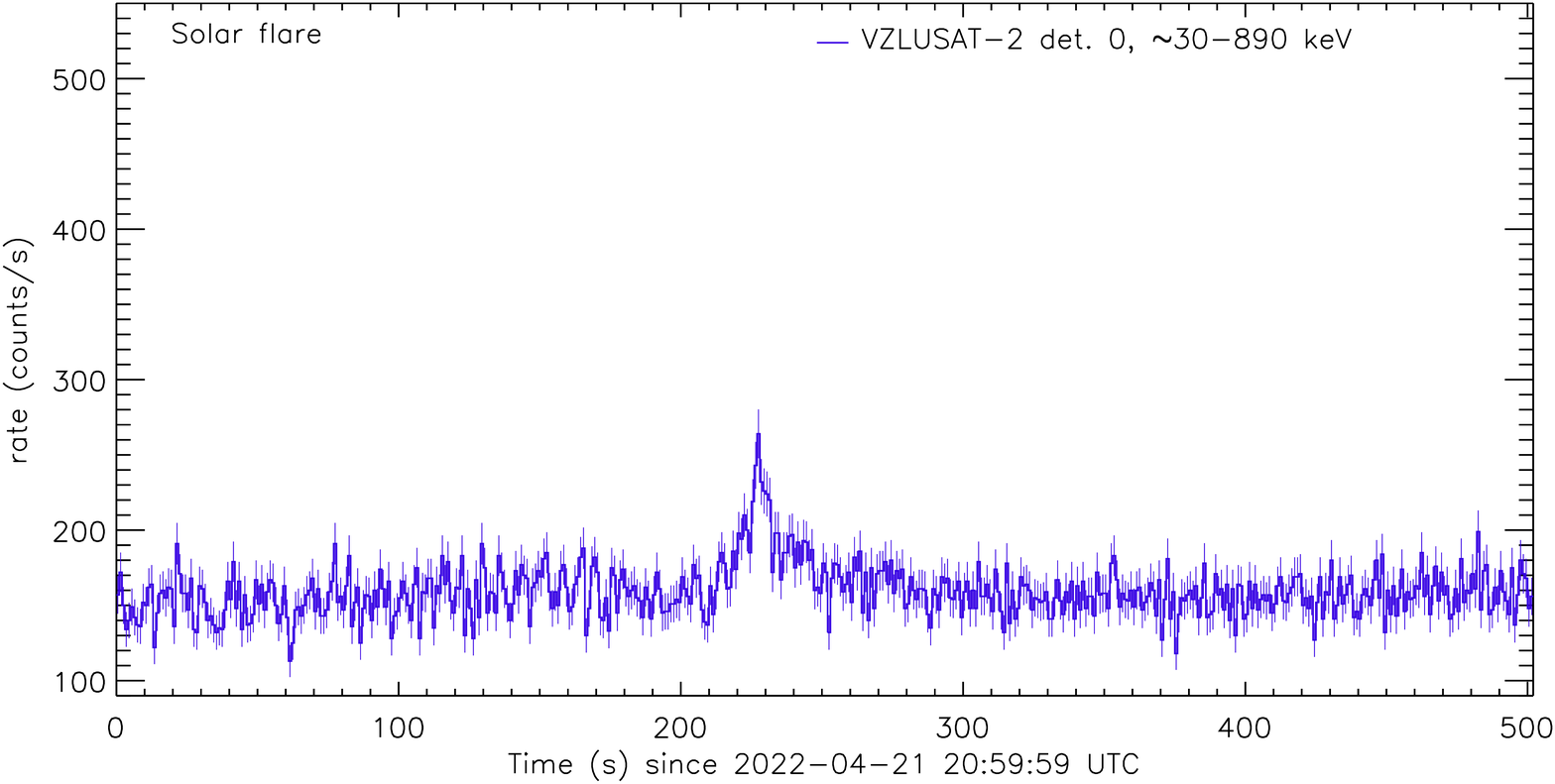}}\vspace*{2mm}
\resizebox{0.49\textwidth}{!}{\includegraphics{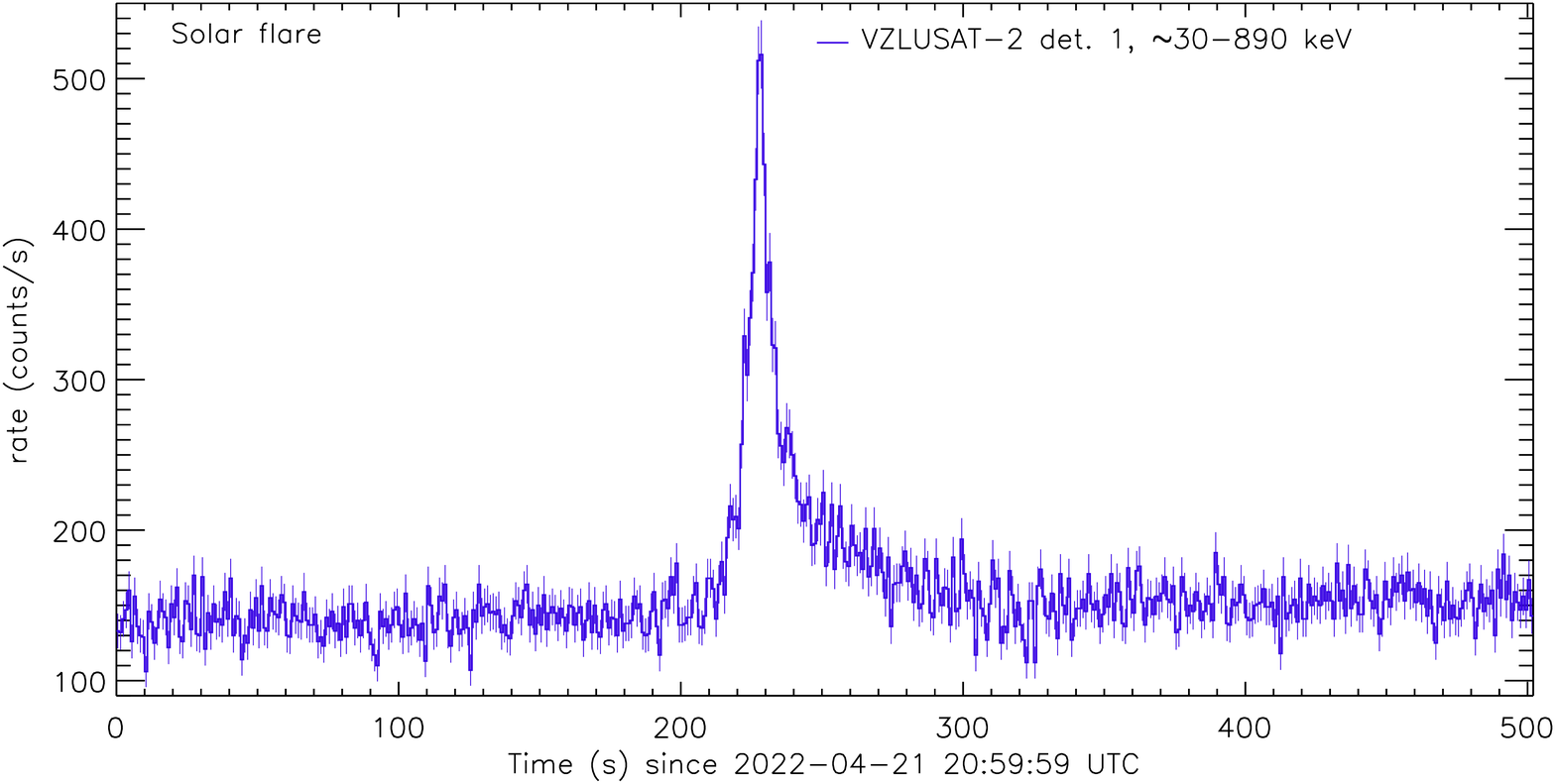}}\vspace*{2mm}
\resizebox{0.49\textwidth}{!}{\includegraphics{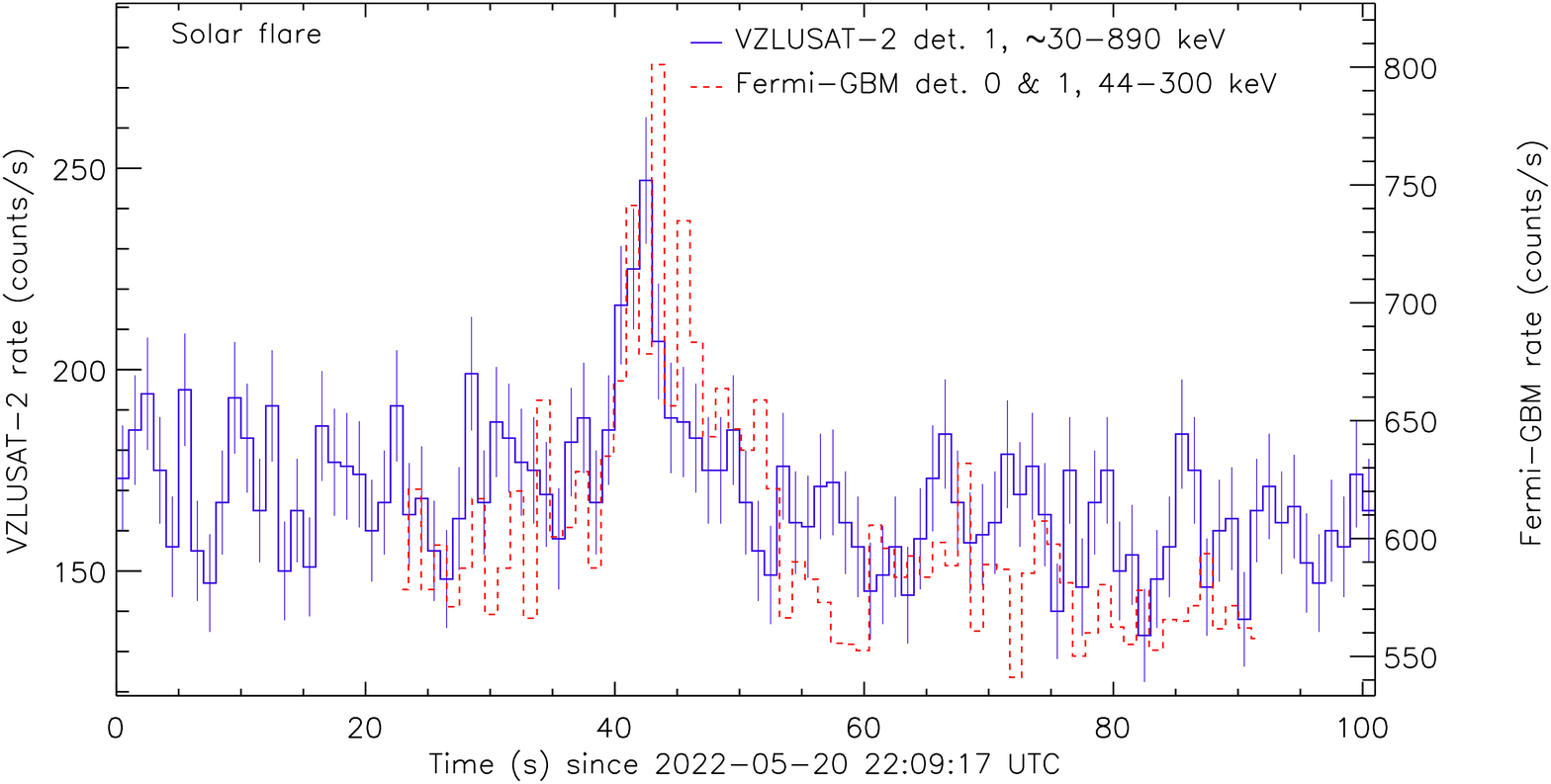}}\vspace*{2mm}
\caption{Two solar flares which occurred on 2022-04-21 and on 2022-05-20 detected by the GRB payload on \textit{VZLUSAT-2}. Top panels show the first solar flare observed by the two perpendicularly placed detector units. The bottom panel shows the second flare over-plotted with the raw light curve measured by \textit{Fermi}/GBM.}
\label{fig:vzlusat2-sol_flares}
\end{center}
\end{figure}

The GRB payload on \textit{VZLUSAT-2} has also detected two Solar flares which coincide with peaks in the X-ray flux measured by GOES-16 and GOES-17\footnote{\url{https://www.swpc.noaa.gov/products/goes-x-ray-flux}}:
on 2022-04-21 (GCN 31937, GCN 31949)\cite{GCN31937,GCN31949} and on 2022-05-20. The observed Solar flares are shown in Fig.~\ref{fig:vzlusat2-sol_flares}.

\subsection{Low Earth Orbit Background Monitoring}
\label{sec:vzlusat2-bkg}

\begin{figure}[p]
\begin{center}
\resizebox{0.49\textwidth}{!}{\includegraphics{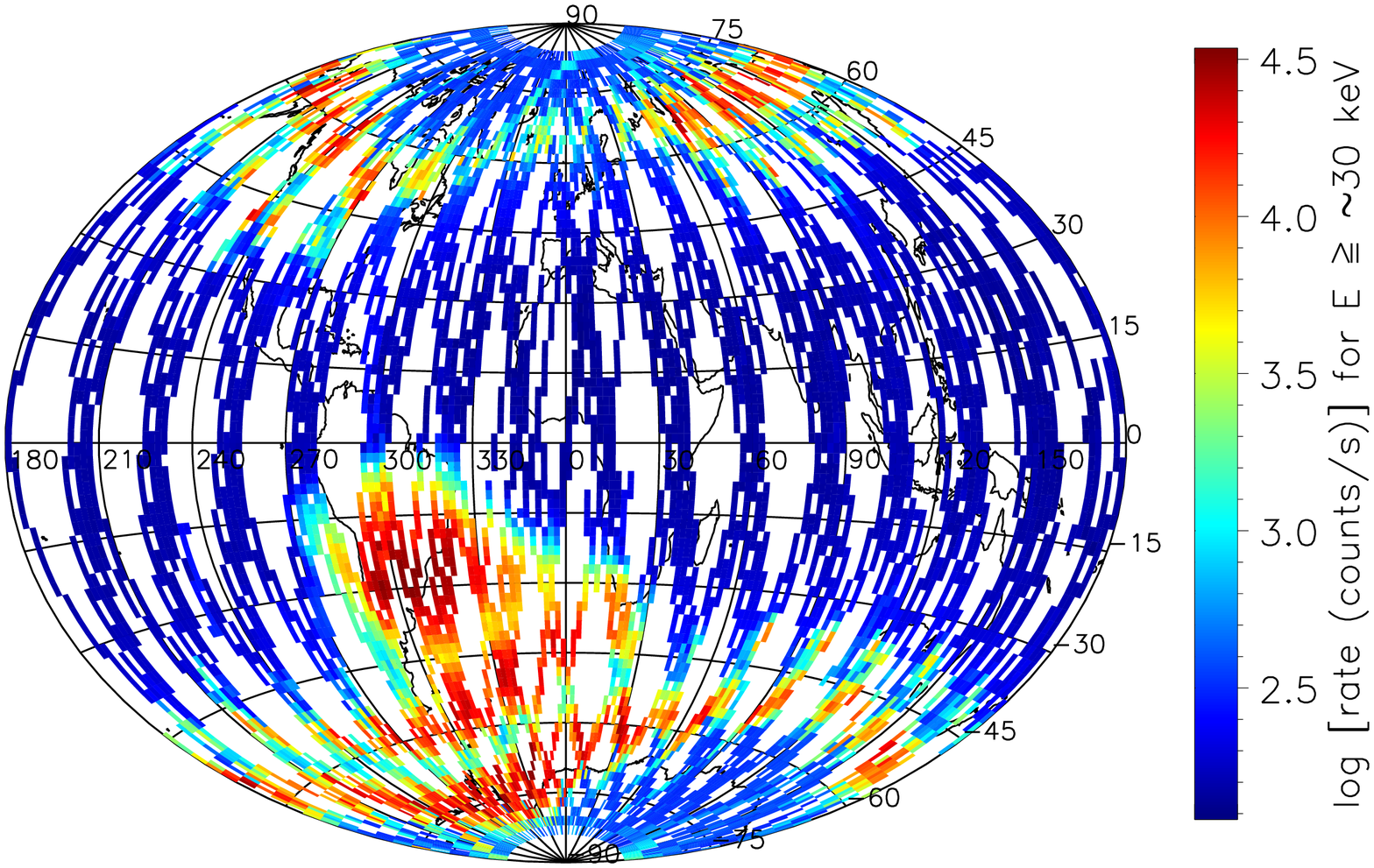}}\vspace*{2mm}
\resizebox{0.49\textwidth}{!}{\includegraphics{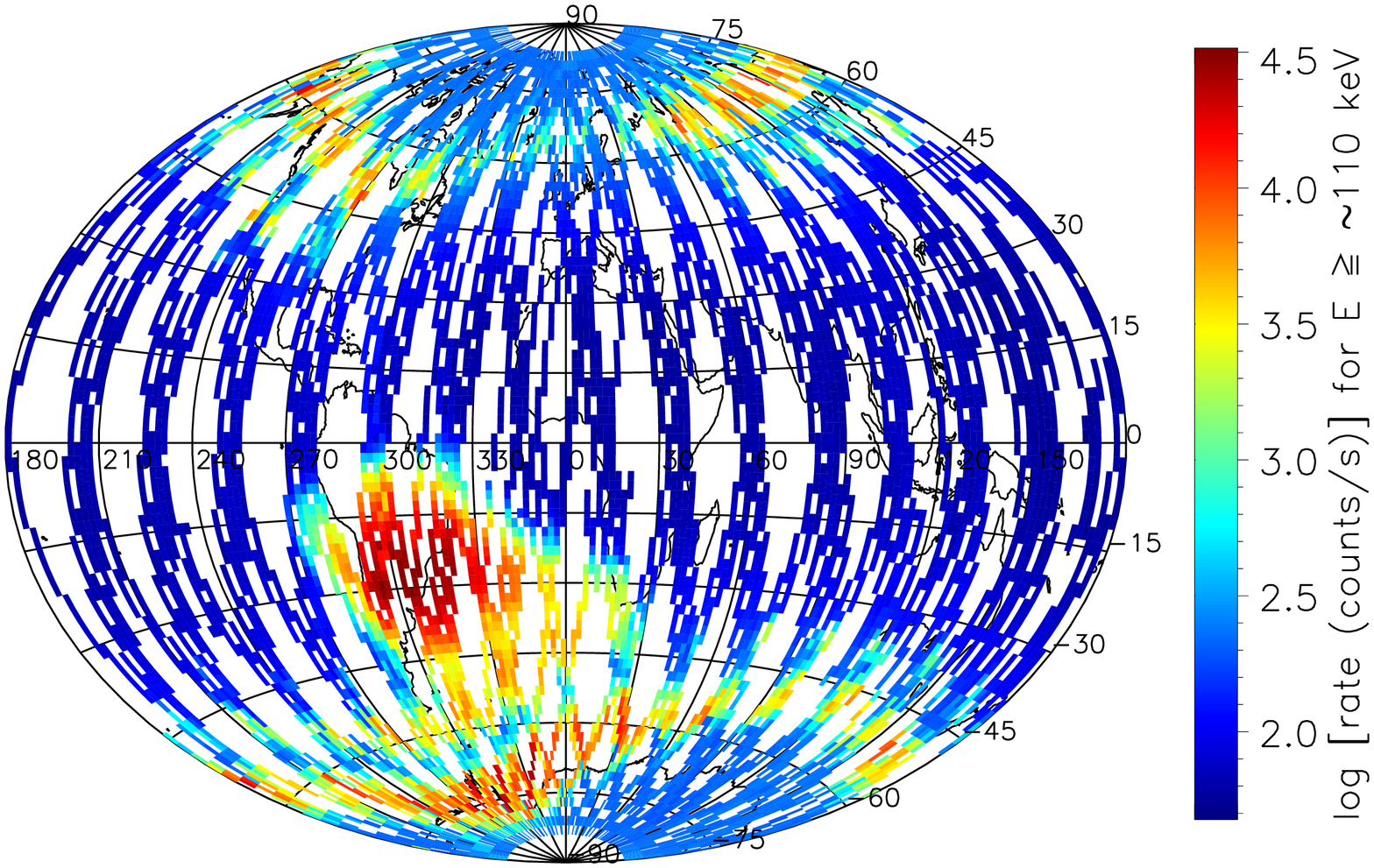}}\vspace*{2mm}
\resizebox{0.49\textwidth}{!}{\includegraphics{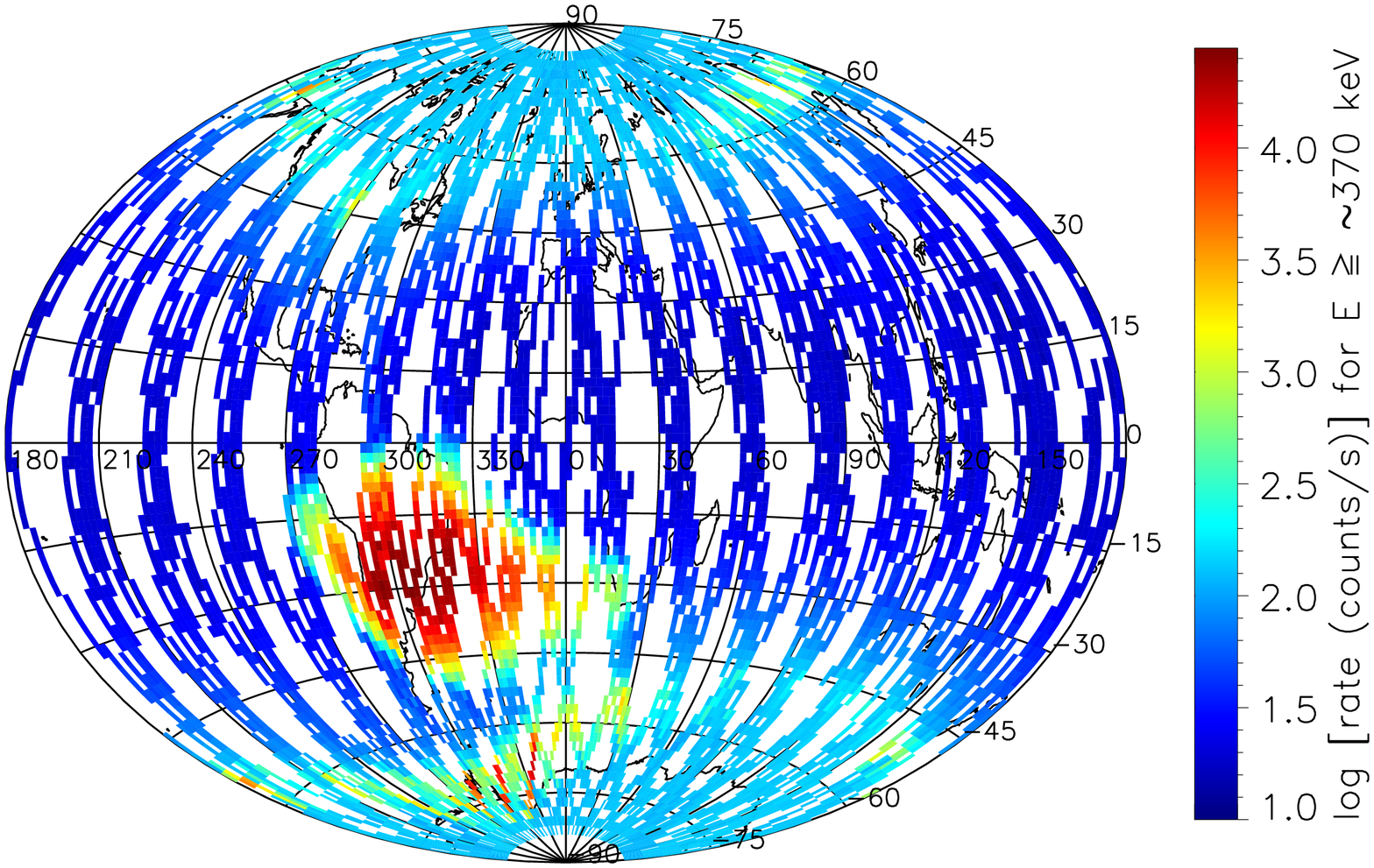}}\vspace*{2mm}
\caption{The background maps at low Earth orbit (520-570\,km altitude) as measured by the GRB payload (detector unit no. 1) on \textit{VZLUSAT-2} for $E\gtrsim 30$\,keV (upper left panel), $E\gtrsim 110$\,keV (upper right panel) and $E\gtrsim 370$\,keV (bottom panel), respectively. The count rate is averaged in a $2^\circ\times2^\circ$ pixels grid.}
\label{fig:vzlusat2-bkg}
\end{center}
\end{figure}

Using the GRB payload on \textit{VZLUSAT-2}, we also construct background maps at LEO. Fig.~\ref{fig:vzlusat2-bkg} shows averaged background count rate maps in different energy ranges collected by the detector unit no. 1.

\begin{figure}[p]
\begin{center}
\resizebox{0.99\textwidth}{!}{\includegraphics{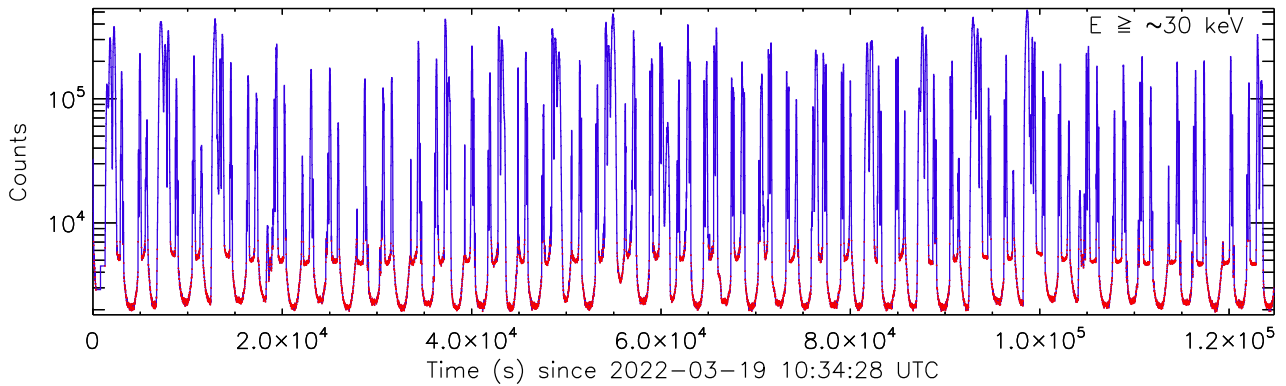}}\vspace*{2mm}
\caption{The background variation at low Earth orbit (520-570\,km altitude) as measured by the GRB payload (detector unit no. 1) on \textit{VZLUSAT-2} for $E\gtrsim 30$\,keV over about one and half day. The number of counts is in 15\,s wide bins. The red points mark the region where a median-intensity long GRB would be detectable with a signal-to-noise ratio SNR=5. This determines the background driven duty cycle.}
\label{fig:vzlusat2-duty_cycle}
\end{center}
\end{figure}

Fig.~\ref{fig:vzlusat2-duty_cycle} shows the background variation at low Earth orbit (520-570\,km altitude) as measured by the GRB payload (detector unit no. 1) on \textit{VZLUSAT-2} for $E\gtrsim 30$\,keV over about one and half day with 15\,s temporal resolution bins. The peaks in the measured background correspond to SAA and polar region passes where the background is dominated by trapped protons and electrons, respectively.

To estimate the fraction of time usable for long GRB detection at a low-Earth SSO, at an altitude of 550\,km we consider a median-intensity long GRB (integrated photon flux above 30\,keV is 1.19\,ph\,cm$^{-2}$\,s$^{-1}$) based on the \textit{Fermi}/GBM FERMIGBRST catalog\footnote{\url{https://heasarc.gsfc.nasa.gov/W3Browse/fermi/fermigbrst.html}} (for details see Ref.~\cite{galgoczi2021}). Next, using Geant4 Monte Carlo simulations as described in Ref.~\cite{galgoczi2021}, the detector's effective area of $\sim 50$\,cm$^2$ and the best incident angle of a median intensity long GRB, we obtain a detection count rate of 0.58\,cnt\,cm$^{-2}$\,s$^{-1}$ for energies $E>30$\,keV. From this value and from the measured background count rate, we obtain a fraction of time of 67\,\% usable for the detection of a typical long GRB if the SNR=5 within 15\,s integration time is required to claim a detection. Red points in Fig.~\ref{fig:vzlusat2-duty_cycle} show the region where a median-intensity long GRB is detectable with the SNR=5.

\section{Summary}
\label{sec:summary}

The \textit{GRBAlpha} and \textit{VZLUSAT-2} CubeSats are successfully operating at LEO and collecting scientific data. \textit{GRBAlpha} has so far detected five GRBs, \textit{VZLUSAT-2} has so far detected three GRBs and two Solar flares demonstrating that nanosatellites can host payloads sensitive enough to routinely detect gamma-ray bursts. Both satellites are monitoring background at LEO and the maps constructed from these measurements will be useful for the preparation of future GRB missions. The degradation of MPPC sensors on {\it GRBAlpha} is being monitored, providing valuable information for future missions which plan to use detectors with MPPC readout. One year after its launch, the \textit{GRBAlpha} detector performance is good and the degradation of the silicon photon counters remains at an acceptable level.

\acknowledgments

We acknowledge support by the grants KEP2/2020 and SA-40/2021 of the Hungarian Academy of Sciences and E\"{o}tv\"{o}s Lor\'{a}nd Research Network, respectively, for satellite components and payload developments and the grant IF-7/2020 for providing the financial support for ground infrastructure. We acknowledge support by the MUNI Award for Science and Humanities funded by the Grant Agency of Masaryk University and by the European Union’s Horizon 2020 Programme under the AHEAD2020 project (grant agreement n. 871158). This work was supported by JSPS KAKENHI Grant Number 21KK0051, JSPS and HAS under Japan - Hungary Research Cooperative Program, and ISAS/JAXA Grant. We are grateful for the operators and developers of the SatNOGS network, helping us with the reception of the first ``sings of life'' telemetry messages as well as providing a nice framework for bulk data download from our detectors flying onboard these satellites. We are also grateful to the community of Japanese radio amateurs, including Tetsurou Sato and Masahiro Sanada, who provided support for the GRBAlpha mission.

\newpage

\bibliography{report} 
\bibliographystyle{spiebib} 

\end{document}